# Coercive fields in ultrathin BaTiO$_3$ capacitors

J. Y. Jo, Y. S. Kim, and T. W. Noh

*ReCOE & FPRD, Department of Physics and Astronomy, Seoul National University, Seoul 151-747, Korea*

Jong-Gul Yoon[a)]

*Department of Physics, University of Suwon, Suwon, Gyunggi-do 445-743, Korea*

T. K. Song

*School of Nano & Advanced Materials Engineering, Changwon National University, Changwon, Gyeongnam 641-773, Korea*

Thickness-dependence of coercive field ($E_C$) was investigated in ultrathin BaTiO$_3$ capacitors with thicknesses ($d$) between 30 and 5 nm. The $E_C$ appears nearly independent of $d$ below 15 nm, and decreases slowly as $d$ increases above 15 nm. This behavior cannot be explained by extrinsic effects, such as interfacial passive layers or strain relaxation, nor by homogeneous domain models. Based on domain nuclei formation model, the observed $E_C$ behavior is explainable via a quantitative level. A crossover of domain shape from a half-prolate spheroid to a cylinder is also suggested at $d \sim$ 15 nm, exhibiting good agreement with experimental results.

In ferroelectric (FE) materials, the direction of polarization ($P$) can be changed by an external electric field ($E$) called a coercive field ($E_C$).[1] Indeed, this physical phenomenon is very important scientifically, as well as technologically. From a technological perspective, most modern devices require low operating voltages to reduce power consumption. In most devices with a FE layer of thickness $d$, the operating voltages should be much larger than $E_C \cdot d$. From a scientific perspective, the $E_C$ can provide valuable information pertaining to polarization switching dynamics. The dependence of $E_C$ on $d$ in thin FE capacitors has been a very important, yet controversial, issue.[1-12]

On the $d$-dependence of $E_C$, extrinsic effects, such as interfacial passive layers and strain relaxation, have been studied extensively. In FE capacitors, interfacial passive layers can be formed between the FE film and conducting electrodes. These detrimental layers induce a significant voltage drop.[2,3] As $d$ decreases, $E_C$ increases: $E_C \propto d^{-1}$.[2-6] Due to the lattice mismatch between the film and substrate, strain may develop inside the FE film. However, as $d$ becomes thicker, the strain relaxes. This relaxation effect typically manifests itself in a shallowing of the FE double-well potential, and results in a smaller $E_C$ value.[7] These extrinsic effects place important technological limits on typical FE device performance.[2-7]

Even for high-quality FE thin films, the $d$-dependence of $E_C$ remains poorly understood. The Landau-Devonshire (LD) theory of ferroelectricity predicts the intrinsic $E_C$, and assumes that the domain switching occurs homogeneously.[8] However, the observed $E_C$ value is always much smaller as a result of FE domain nucleation.[8] Kay and Dunn studied the $d$-dependence of $E_C$ in the context of the nucleation model.[11,12] This model predicts a semi-empirical scaling law, where $E_C \propto d^{-2/3}$.[12] Recently, Ducharme *et al.* investigated ultrathin films of a vinylidene fluoride copolymer as thin as 1 nm and claimed that the measured $E_C$ values below 15 nm are independent of $d$, which agreed with the LD predictions (*i.e.*, the thermodynamically intrinsic value).[8] However, several workers later pointed out that the observed $E_C$ could not be thermodynamically intrinsic.[9,10] The possibility of reaching the intrinsic $E_C$ value in ultrathin FE films remains controversial.[8-10]

In this letter, we report the $d$-dependence of $E_C$ using high-quality BaTiO$_3$ (BTO) capacitors and compare our results with theoretical predictions. The fabrication of fully-strained SrRuO$_3$(SRO)/BTO/SRO heterostructures has been reported recently, with $d$ values between 5 and 30 nm, on SrTiO$_3$ (001) substrates.[13] We measured their $P$-$E$ hysteresis loops with frequencies from 2 to 100 kHz. Figure 1 shows the $P$-$E$ hysteresis loops of the BTO capacitors measured at 100 kHz. The measured $E_C$ value with increasing $E$-field ($|E_C^+|$) is slightly different when compared with the decreasing $E$-field ($|E_C^-|$) results. Due to this "imprint phenomena", the experimental $E_C$ values were determined using $E_C = (|E_C^+| + |E_C^-|)/2$. The $E_C$ values at 2 and 100 kHz are shown in Fig. 2 as solid squares and open circles, respectively. Their significant variation with the frequency change indicates that the domain dynamics play an important role in determining the $E_C$ values.

Earlier accounts of ultrathin FE capacitors reported that $E_C$ increases drastically as $d$ decreases: typically, $E_C \propto d^{-1}$.[2-7] By contrast, in our $P$-$E$ hysteresis loops, we observed a rather weak dependence of $d$ with respect to $E_C$. Due to the high-quality BTO capacitors used, we can confidently eliminate the extrinsic effects.[13,14] Note that the observed weak $d$-dependent $E_C$ behavior is quite different from the $d^{-1}$ behavior predicted by the interfacial passive layer model.[2-4] In addition, our observations are consistent with a recent report that demonstrated that the capacitors were nearly free from interfacial passive layers.[14] Our films were almost fully strained,[13] and therefore strain relaxation could not play an important role in the $E_C$ behavior. Consequently, the high quality of the BTO capacitors used afforded us with a great opportunity to investigate the intrinsic behavior of $E_C$ in ultrathin FE films.[13,14]

At this point, let us compare the observed $E_C$ values with the thermodynamic intrinsic limit using the LD model. To check for the possibility of homogeneous switching in our ultrathin films, we calculated $E_C$ using LD theory with the free energy density functional ($\hat{G}$):

$$\hat{G} = a_1 P^2 + a_{11} P^4 + a_{111} P^6 - EP, \qquad (1)$$

where $a_1$ and $a_{11}$ are functions of strain.[15] We can estimate $E_C$ by minimizing $\hat{G}$.[7] The red dashed line in Fig. 2 shows the LD calculation results for $E_C$, where the BTO films are assumed to be unstrained. It is already well known that a compressive strain in an FE layer may result in a higher $P$ value.[15] Since our BTO films are under such compressive strains, their $E_C$ values may in fact be higher. To determine the effects of the compressive strains, we reapplied Eq. (1) with modified values of $a_1$ and $a_{11}$, following the work by Pertsev *et al.*[15] With a compressive strain of 2.2%, we obtained the thermodynamic intrinsic $E_C$ values in the strained state. As depicted by the black dotted line in Fig. 2, *the LD theoretical $E_C$ values are larger than our experimental results by a factor of approximately 5.*

In the FE capacitor geometry, the polarization charges are screened by free carriers inside the electrode. However, the screening is incomplete due to the finite screening length of the free carriers, and results in a depolarization field ($E_d$). Numerous recent reports have highlighted how $E_d$ plays a very important role in the physical properties of ultrathin FE capacitors. Therefore, it is useful to identify the effects of $E_d$ on the homogeneous domain switching. From the electrostatic consideration under the short-circuit condition of a capacitor,[16] $E_d$ can be calculated as

$$E_d = -P(2\varepsilon_F / d) / \varepsilon_0 \varepsilon_F (2\varepsilon_F / d + \varepsilon_e / \lambda) \equiv -P(1-\theta)/\varepsilon_0 \varepsilon_F, \qquad (2)$$

where $\varepsilon_e$ ($\varepsilon_F$) is the dielectric constant of the electrode (FE layer), $\lambda$ is the screening length of the electrode material, and $\theta$ is the compensation ratio for polarization charge. With the contribution of the depolarization field term (*i.e.*, $-E_d \cdot P$), $\hat{G}$ can be rewritten as

$$\hat{G} = \left(a_1 + (1-\theta)/2\varepsilon_0\varepsilon_F\right)P^2 + a_{11}P^4 + a_{111}P^6 - EP. \tag{3}$$

Since $a_1$ (a negative value) determines the stability of the ferroelectricity,[15] the positive depolarization contribution of $(1-\theta)/2\varepsilon_0\varepsilon_F$ helps to suppress $P$ and $E_C$. By minimizing $\hat{G}$ in Eq. (3), we can estimate the $d$-dependent $E_C$. The blue solid line in Fig. 2 shows the $d$-dependent $E_C$ values using the parameter values extracted from Ref. [17]. Indeed, as $d$ decreases, the calculated $E_C$ value decreases rapidly. This indicates that the $E_d$ can help domain switching, especially in the ultrathin film region. Our calculations demonstrate that the homogenous switching models cannot explain the experimental data for ultrathin BTO capacitors.

In most FE materials, domain switching occurs inhomogeneously by forming domain nuclei and propagating reversed domains.[18] Figure 3(a) presents a schematic diagram for the formation of half-prolate spheroidal nuclei with a reversed polarization direction. From the electrostatic energy calculations, the nucleation energy ($U$) can be defined as the sum of the external field contribution, domain-wall energy, and depolarizing energy ($U_d$) between the nucleus and surrounding domain:[11,12]

$$\begin{aligned} U &= -2PEv + \sigma_w s + U_d \\ &= -ar^2 l + brl + cr^4/l, \end{aligned} \tag{4}$$

where $v$, $s$, $r$, and $l$ represent the volume, area, radius, and length of the nucleus, respectively. Note that $a = 4\pi PE/3$, $b = \pi^2\sigma_w/2$, and $c = 4\pi P^2/3\varepsilon_0\varepsilon_a(\ln(2l/r(\varepsilon_c/\varepsilon_a)^{1/2}) - 1)$, where $\varepsilon_a$ ($\varepsilon_c$) is the $a$ ($c$)-axis dielectric constant of the FE layer. The term $\sigma_w$ is the domain wall energy per area. Due to the competition between the terms $-2PEv$ and $U_d$, $U$ has a saddle point at a finite value of $l$, denoted by $l^*$. This $l^*$ represents the characteristic size of a stable domain nucleus, called the critical length. It is possible to obtain $E_C$ values from the saddle points in $U$ in $(r, l)$-space using measured $P$ values,[17] $\varepsilon_c \approx 80$,[17] and $\varepsilon_a \approx 220$.[15,19] We used the value of $\sigma_w$ as the only free parameter. The red solid line in Fig. 3(c) shows the calculated $E_C$ values when $\sigma_w = 17$ mJ/m$^2$.

Note that this domain wall energy value agrees with the theoretical values,[20] and also predicts $l^*$ ≈ 15 nm for the experimental $E_C$. In addition, the calculated $E_C$ values are proportional to $d^{-2/3}$, which is close to the experimental values (*i.e.*, solid square) in the region of $d$ = 15 ~ 30 nm.

When the film becomes very thin, $d$ can be smaller than $l^*$. Therefore, the stable nucleus shape should be a cylinder instead of a half-prolate spheroid, as depicted schematically in Fig. 3(b). As a result, term $U_d$ in Eq. (4) vanishes, and $a = 2\pi PE$ and $b = 2\pi\sigma_w$. Following a similar surmounting process for $U$, $E_C$ can be independent of $d$.[11,12] The green dashed line in Fig. 3(c) illustrates the calculated $E_C$ values for the cylindrical nuclei when $\sigma_w$ = 17 mJ/m$^2$ is used. Note that the green dotted line highlights the good agreement with the experimental $E_C$ values for $d \leq 15$ nm. Due to two different domain nucleation processes, the $d$-dependence of $E_C$ shows crossover behavior at $d \approx l^* \approx 15$ nm. The electrostatic $E_C$ calculation based on two domain nucleation shapes provides a reasonable explanation for our experimental $E_C$ results.

Nevertheless, the nucleation process in most FE materials is known to be impossible thermodynamically, which is commonly referred to as "Landauer's paradox".[11] Landauer emphasized that the thermodynamic nucleation process requires an insurmountable energy barrier ($U^*$) compared to the available thermal energy.[11] However, we recently showed that, in our ultrathin BTO FE film, the huge $E_d$ can make $U^* \sim 10\ k_BT$.[21] This is the important reason why the $d$-dependent $E_C$ behavior can be explained in terms of the domain nuclei formation model.

Finally, it is valuable to look at the technological meaning of our observations. As we mentioned in the introductory part, the coercive voltage ($V_C$) is a crucial parameter for device applications, since it is related to the device operating voltage directly. In our high-quality ultrathin BTO capacitors, $V_C$ decreases almost linearly with $d$, as demonstrated in the inset of Fig.

3(c). This suggests that, for FE capacitors without interfacial passive layers, the operating voltage limitation due to $V_C$ must not be a major obstacle for device integration.

In summary, we investigated the thickness-dependence of the coercive field in $SrRuO_3/BaTiO_3/SrRuO_3$ capacitors with a $BaTiO_3$ thickness between 5 and 30 nm. We showed that homogenous polarization switching cannot occur in our high-quality capacitors. Based on the domain nuclei formation model, we can explain the thickness-dependent coercive fields of ultrathin $BaTiO_3$ capacitors quantitatively.

This work was supported financially by Creative Research Initiatives (Functionally Integrated Oxide Heterostructure) of MOST/KOSEF.

[a] e-mail : jgyoon@suwon.ac.kr

**References**

[1] J. F. Scott, *Ferroelectric Memories* (Springer-Verlag, New York, 2000).

[2] J. F. M. Cillessen, M. W. J. Prins, and R. W. Wolf, J. Appl. Phys. **81**, 2777 (1997).

[3] P. K. Larsen, G. J. M. Dormans, D. J. Taylor, and P. J. van Veldhoven, J. Appl. Phys. **76**, 2405 (1994).

[4] A. K. Tagantsev and I. A. Stolichnov, Appl. Phys. Lett. **74**, 1326 (1999).

[5] A. K. Tagantsev, M. Landivar, E. Colla, and N. Setter, J. Appl. Phys. **78**, 2623 (1995).

[6] N. Yanase, K. Abe, N. Fukushima, and T. Kawakubo, Jpn. J. Appl. Phys. **38**, 5305 (1999).

[7] N. A. Pertsev, J. Rodríguez Contréras, V. G. Kukhar, B. Hermanns, H. Kohlstedt, and R. Waser, Appl. Phys. Lett. **83**, 3356 (2003).

[8] S. Ducharme, V. M. Fridkin, A.V. Bune, S. P. Palto, L. M. Blinov, N. N. Petukhova, and S. G. Yudin, Phys. Rev. Lett. **84**, 175 (2000).

[9] A. M. Bratkovsky and A. P. Levanyuk, Phys. Rev. Lett. **87**, 019701 (2001).

[10] R. L. Moreira, Phys. Rev. Lett. **88**, 179701 (2002).

[11] R. Landauer, J. Appl. Phys. **28**, 227 (1957).

[12] H. F. Kay and J. W. Dunn, Phil. Mag. **7**, 2027 (1962).

[13] Y. S. Kim *et al.*, Appl. Phys. Lett. **86**, 102907 (2005).

[14] Y. S. Kim *et al.*, Appl. Phys. Lett. **88**, 072909 (2006).

[15] N. A. Pertsev, A. G. Zembilgotov, and A. K. Tagantsev, Phys. Rev. Lett. **80**, 1988 (1998).

[16] R. Mehta, B. D. Silverman, and J. T. Jacobs, J. Appl. Phys. **44**, 3379 (1973).

[17] D. J. Kim *et al.*, Phys. Rev. Lett. **95**, 237602 (2005).

[18] T. Tybell, P. Paruch, T. Giamarchi, and J.-M. Triscone, Phys. Rev. Lett. **89**, 097601 (2002).

[19] Z.-G. Ban and S. P. Alpay, J. Appl. Phys. **91**, 9288 (2002).

[20] J. Padilla, W. Zhong, and D. Vanderbilt, Phys. Rev. B **53**, R5969 (1996).

[21] J. Y. Jo *et al.*, preprint at <http:// arvig.org/cond-mat/0605079>.

**Figure captions**

FIG. 1. (Color online) (a) Polarization-external electric field (*P-E*) hysteresis loops of SrRuO$_3$/BaTiO$_3$/SrRuO$_3$ capacitors with BaTiO$_3$ thickness (*d*) from 30 to 5.0 nm.

FIG. 2. (Color online) Coercive fields ($E_C$) based on homogeneous polarization switching models. Calculated intrinsic $E_C$ of BaTiO$_3$ capacitors with the Landau-Devonshire theory and measured $E_C$ values (open circles) at 100 kHz and (solid squares) at 2 kHz. The red dashed line and black dotted line are the calculated $E_C$ for unstrained and −2.2% fully strained BaTiO$_3$ bulk, respectively. The blue solid line is the calculated $E_C$ under −2.2% fully strain and a depolarization field.

FIG. 3. (Color online) Schematic diagrams of (a) half-prolate spheroidal nucleation and (b) cylindrical nucleation. (c) The *d*-dependent (solid squares) experimental and calculated $E_C$ based on the half-prolate spheroidal nucleation model (red solid line) and the cylindrical nucleation model (green dotted line). In the shaded region, the *d*-dependent $E_C$ is thought to be governed by cylindrical nucleation, rather than half-prolate spheroidal nucleation. The inset shows the coercive voltages ($V_C$) as a function of *d*. The black dashed line is an eye-guide.

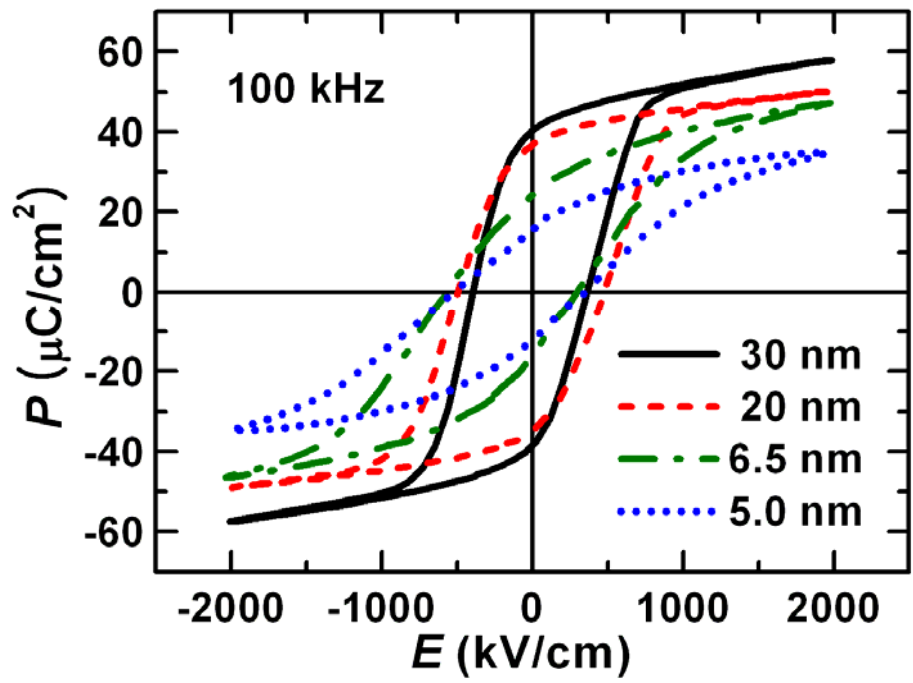

**Figure 1**

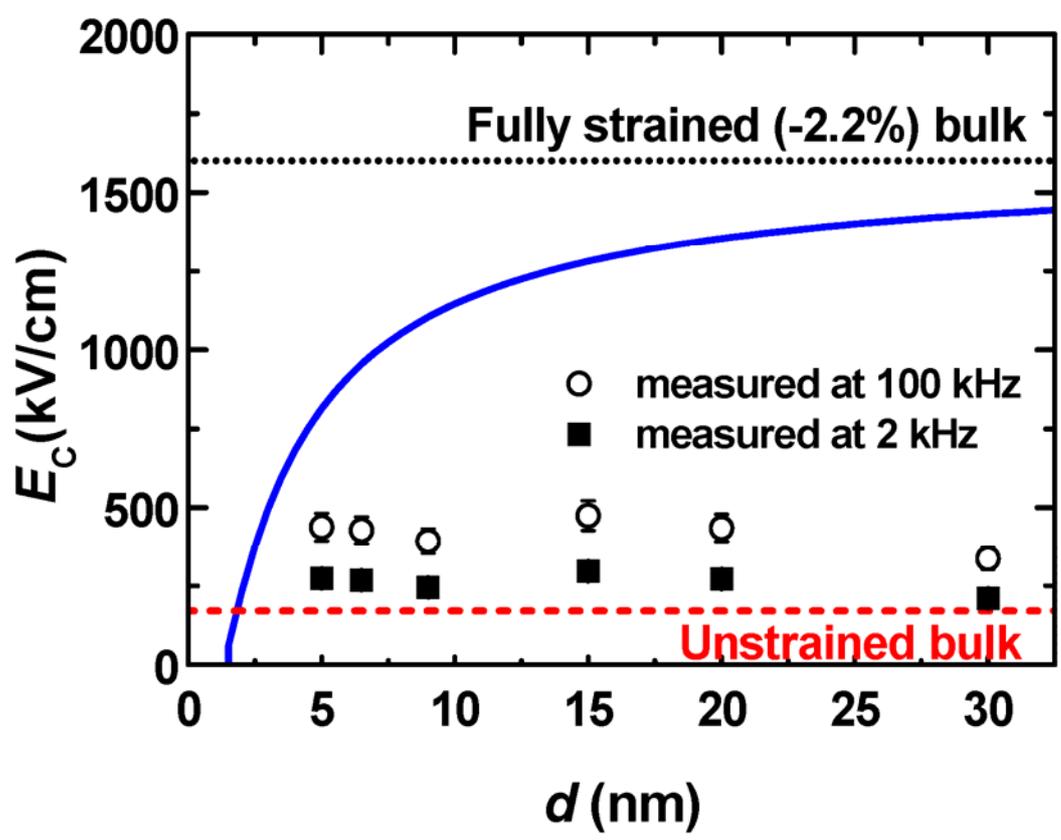

**Figure 2**

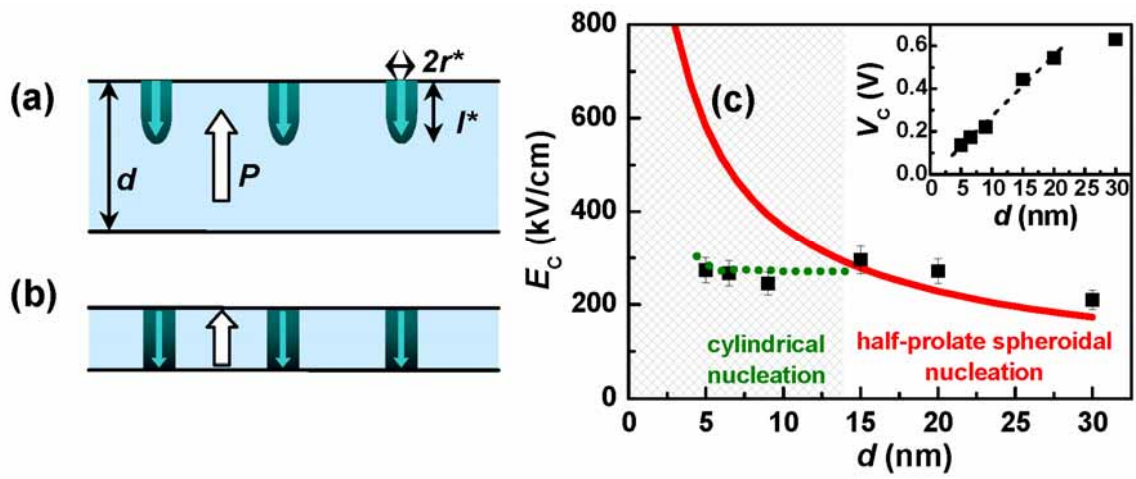

**Figure 3**